\documentclass[fleqn,usenatbib]{rasti}

\usepackage{newtxtext,newtxmath}
\usepackage{mathtools}

\usepackage[T1]{fontenc}

\DeclareRobustCommand{\VAN}[3]{#2}
\let\VANthebibliography\thebibliography
\def\thebibliography{\DeclareRobustCommand{\VAN}[3]{##3}\VANthebibliography}

\usepackage{graphicx}	
\usepackage{amsmath}	

\usepackage{enumitem}
\usepackage{array,multirow}
\usepackage[caption=true]{subfig}

\title[Interferometric Imaging with Closure Invariants]{Interferometric Image Reconstruction using Closure Invariants and Machine Learning}

\author[N. Thyagarajan et al.]{
Nithyanandan Thyagarajan,$^{1}$\thanks{E-mail: Nithyanandan.Thyagarajan@csiro.au}
Lucas Hoefs,$^{1,2}$
and O. Ivy Wong$^{1,3}$
\\
$^{1}$Commonwealth Scientific and Industrial Research Organisation (CSIRO), Space \& Astronomy, P. O. Box 1130, Bentley, WA 6102, Australia\\
$^{2}$Mechatronics Engineering Department,  Curtin University, Bentley, WA 6102, Australia\\
$^{3}$International Centre for Radio Astronomy Research, The University of Western Australia, Crawley, WA 6009, Australia
}

\date{Accepted 2024 August 1. Received 2024 July 19; in original form 2023 November 9}

\pubyear{2024}

\begin{document}
\label{firstpage}
\pagerange{\pageref{firstpage}--\pageref{lastpage}}
\maketitle

\begin{abstract}
  Interferometric closure invariants encode calibration-independent details of an object's morphology. Excepting simple cases, a direct backward transformation from closure invariants to morphologies is not well established. We demonstrate using simple Machine Learning models that closure invariants can aid in morphological classification and parameter estimation. We consider six phenomenologically parametrised morphologies: point-like, uniform circular disc, crescent, dual disc, crescent with elliptical accretion disc, and crescent with double jet lobes. Using logistic regression (LR), multi-layer perceptron (MLP), and random forest models on closure invariants obtained from a sparsely covered aperture, we find that all methods except LR can classify morphologies with $\gtrsim$80\% accuracy, which improves with greater aperture coverage. Separately from the classification problem, given an independently confirmed class, we estimate parameters of uniform circular disc, crescent, and dual disc morphologies using simple MLP models, and parametrically reconstruct images. The estimated parameters and images correspond well with inputs, but the accuracy worsens when degeneracies between parameters are present. This independent approach to interferometric imaging under challenging observing conditions such as that faced by the Event Horizon Telescope and Very Long Baseline Interferometry in general can complement other methods in robustly constraining an object's morphology.
\end{abstract}




\section{Introduction} \label{sec:intro}

Image synthesis using radio interferometric measurements requires an array of receiver elements sampling spatial correlations of the radiation incident on the aperture to infer the spatial intensity distribution on the sky within the telescope's field of view \citep{TMS2017,SIRA-II}. As the spatial resolution of the image scales inversely with the largest spacing between the interferometer array elements, obtaining very fine details in the image requires a technique called Very Long Baseline Interferometry (VLBI) which requires array elements widely separated from each other, with typical separations spanning continental or even planet-sized scales. Due to the sparseness of measurements and demanding requirements of accurate signal calibration required in maintaining a high degree of spatial coherence while combining signals from such large separations, VLBI imaging is extremely challenging in general \citep{TMS2017,VLBI-SIRA}. 

There are certain invariants in interferometry like closure phases \citep{Jennison1958} and closure amplitudes \citep{Twiss+1960} that are immune to propagation and instrumental effects that are associable with individual array elements, and are thus independent of calibration and errors therein. Being true observables of the observed object's morphology, they implicitly or explicitly serve as useful anchors for inferring an object's morphology. A few examples of classic VLBI successes that used closure quantities include the discovery of the double-lobed structures of Cygnus~A \citep{Jennison1957,Jennison+Latham1959} and Centaurus~A \citep{Twiss+1960}, determination of the core-jet morphology of quasar 3C~147 \citep{Wilkinson+1977}, and providing the first direct evidence for superluminal expansion of the relativistic jet in quasar 3C~273 \citep{Pearson+1981}. A recent example is the event horizon-scale imaging using the Event Horizon Telescope (EHT) data of the central supermassive black holes at the centres of M87 \citep{eht19-1,eht19-2,eht19-3,eht19-4,eht19-5,eht19-6} and Sgr~A$^\star$ \citep{eht21-7} by the EHT collaboration (EHTC).

The sparseness of data, low signal-to-noise conditions, choices of initial models in deconvolution methods, even small miscalibrations, and other unconscious biases in the analysis can lead to artefacts and diverging image morphologies, and thus misinterpretations of the results. For instance, \citet{Carilli+Thyagarajan2022} detailed the effects of the choice of initial models in the iterative imaging process. The choice of introducing unconstrained ``boxes'' in the CLEAN algorithm \citep{Hogbom1974,Schwab1984} can lead to divergent results \citep{Deconvolution-SIRA,Miyoshi+2022}. The EHTC have acknowledged these challenges and followed a very detailed verification process to mitigate these risks. Subsequent analysis by other groups using different methods have found consistent results \citep{Sun+Bouman2021,Carilli+Thyagarajan2022,Broderick+2022,Arras+2022,Medeiros+2023}. Nevertheless, independent confirmation by independent methods is paramount in such important scientific studies. 

Closure invariants, being independent of element-based calibration, have the potential to be largely immune to the aforementioned risks present in traditional interferometric image synthesis. While closure invariants are mathematically well-defined and have been extensively used for decades, their physical interpretation has been unclear. A first geometric understanding of closure phase was presented in \citet{Thyagarajan+Carilli2022}. Closure invariants are adept at distinguishing point-like and centrosymmetric objects (unit closure amplitudes and vanishing closure phases) from non-symmetric morphologies, and corruptions that are element-based rather than spatially correlated \citep{TMS2017}. Similarly, they can be used to infer the presence of polarisation inherent to the object \citep{Broderick+2020b,Samuel+2022}. 

Interferometric imaging based on closure invariants has been studied previously. For example, \citet{Chael+2018} employ a minimisation scheme over closure quantities supplemented by some \textit{a priori} information and \textit{regularisers} that favour certain image features. Probabilistic sampling of the posterior distribution and forward-modeling of the sampled morphological class parameters has been used to determine the best fit to the measured closure invariants \citep{eht19-6,Broderick+2020a,Tiede+2022,Saurabh+Nampallivar2023}. Variational deep probabilistic imaging (DPI) approaches have been proposed \citep{Sun+Bouman2021,Sun+2022} that, without training data, optimise the weights of a neural network to learn the posterior distribution of an unobserved image, from which image samples are generated to fit a particular measurement dataset such as closure invariants. However, it is still notable that unlike the analytical relationship between visibilities and image intensities, a direct backward- or inverse-transformation to the object's morphology from a given set of closure invariants has not been clearly established.

In this work, our primary motivation is to demonstrate a proof-of-concept that closure invariants can be used to distinguish between different image morphologies and provide an independent pathway for parametric image reconstruction using machine learning (ML) methods, wherein propagating the closure invariants through the weights of the layers of the ML model could provide a direct backward-transform to the image morphology. The paper has two independent objectives, namely, to use calibration-independent closure invariants and simple ML models to (1) provide morphological classifications of images, and (2) parametrically reconstruct images. Here, we take a broad theoretical view and approach the classification and parametrisation as independent problems, both of which separately are relevant in interferometric inference of image morphologies. The latter is particularly relevant in situations where parameters have to be estimated when the classification is determined \textit{a priori} such as the presence of FR-II core-jet quasar morphology, a relativistic jet expanding from an active galactic nucleus, a faint planet orbiting a star, asymmetry on a stellar surface, etc.

The paper is organised as follows. We introduce interferometric closure invariants in \S\ref{sec:CI}. We present a proof of concept for morphological classification using simple ML methods in \S\ref{sec:classification}, where we examine the accuracy of our simple ML classifiers across varying degrees of aperture coverage, noise, and morphological complexity, as well as their performance on classifying data from classes not included in training. A proof of concept for estimating the morphological parameters and reconstructing images using ML methods for some chosen morphologies is presented in \S\ref{sec:parametrisation}, where we explore the sensitivity of closure invariants to the morphological parameters and degeneracies between them. \S\ref{sec:summary} presents a discussion and summary. Appendix \ref{sec:ML-graphs} contains supporting material on the ML models used.

\section{Closure Invariants in Interferometry} \label{sec:CI}

The concepts of closure phases \citep{Jennison1958} and closure amplitudes \citep{Twiss+1960} have been in use in radio astronomy for many decades. They have been integral to advances in calibration and interferometric synthesis imaging \citep{Selfcal-SIRA,TMS2017,Carilli+2022}. They were extended to polarimetric measurements through the formalism of closure traces \citep{Broderick+2020b}. Recently, a unified and general theory of interferometric closure invariants for co-polar and polarimetric measurements has emerged from \citet{Thyagarajan+2022} and \citet{Samuel+2022}, respectively. We will primarily use the former because this work pertains to co-polar measurements. Below is a brief outline of the theory of closure invariants in co-polar interferometry relevant for this work. 

In radio interferometry, the basic measurement units are spatial correlations, known as \textit{visibilities}, in the aperture plane corresponding to different array element spacings. In an interferometer array with $N$ elements labelled by the indices $a=0,\ldots,N-1$, each element, $a$, measures the amplitude and phase of the stochastic electric field, $e_a$ (represented by a complex number), incident on it. The true spatial correlation (visibility) between pairs of array elements is $V_{ab}\coloneqq \langle e_a e_b^\dagger \rangle$, where, $\dagger$ denotes complex conjugation. 

The visibilities are related to the spatial distribution of intensities on the sky or image plane \citep{vanCittert34,Zernike38}. Under certain reasonable assumptions \citep{TMS2017,Coherence-SIRA}, these visibilities denote the Fourier components (in the aperture plane) of the images that we are aiming to classify and reconstruct,
\begin{align} 
    V_{ab} \equiv V(\mathbf{u}_{ab}) &= \iint_\mathbb{S} I(\hat{\mathbf{s}}) \, e^{-2\pi i \mathbf{u}_{ab}\cdot \hat{\mathbf{s}}} \, \mathrm{d}^2\hat{\mathbf{s}} \, , \label{eqn:rime}
\end{align}
where, $\hat{\mathbf{s}}$ denotes a unit vector covering the sky surface, $\mathbb{S}$. $\mathbf{u}_{ab}\equiv (u,v)$ denotes the spacial frequency and is related to the spacing between elements, $\mathbf{x}_{ab}$, as $\mathbf{u}_{ab}=\mathbf{x}_{ab}/\lambda$, where, $u$ and $v$ refer to $x$- and $y$-components of $\mathbf{u}_{ab}$ in the aperture (Fourier) plane, and $\lambda$ is the wavelength. $I(\hat{\mathbf{s}})$ is the spatial distribution of intensities on $\mathbb{S}$. 

In real-world measurements, corruptions caused by propagation and instrumental effects affect the array element's measurement as $e_a^\prime=g_a e_a$, where, $g_a$ is a complex corruption factor. Therefore, the measured visibility is also corrupted as $V_{ab}^\prime = g_a V_{ab} g_b^\dagger$. This necessitates calibration and elimination of these multiplicative corrupting factors. There are several situations when accurate calibration can be challenging and can lead to artefacts in the images \citep{Selfcal-SIRA}. 

Invariants like closure phases and amplitudes are special interferometric quantities constructed in specific ways that eliminate the corrupting effects caused by the $g_a$ terms partially or altogether \citep{TMS2017,Selfcal-SIRA}. Because of this property, even when they are constructed with corrupted data, they remain true observables of the physical system under study, independent of local corrupting factors\footnote{Although means of closure invariants are independent of local corrupting factors, their higher order moments are not necessarily so \citep{Blackburn+2020}.}. 

In this work, we adopt the Abelian gauge theory formalism to obtain a complete and independent set of closure invariants in co-polar interferometric measurements \citep[see][for details]{Thyagarajan+2022}. We start by identifying a specific antenna location as the reference vertex (indexed as element 0) and constructing all unique closed triangular loops starting and ending with this vertex \citep{TMS2017}. For an $N$-element array, this gives us $(N-1)(N-2)/2$ independent triangular loops. On each triangular loop, we define an \textit{advariant} as
\begin{align}
    \mathcal{A}_{0ab}^\prime &= V_{0a}^\prime \widehat{V}_{ab}^\prime V_{b0}^\prime = |g_0|^2 \mathcal{A}_{0ab} \, , a\in[1..N-1] \, , a < b \le N-1 \, , \label{eqn:advariant} 
\end{align}
where, $\widehat{Z}=(Z^\dagger)^{-1}$ for a non-zero complex number, $Z$. There are $(N-1)(N-2)/2$ complex advariants (one per independent triangle), each with a pair of real and imaginary parts. This amounts to $(N-1)(N-2)$ real numbers, all with the same unknown scaling factor, $|g_0|^2$, that is associated with the reference vertex. In the final step, the sought closure invariants are obtained from the advariants by eliminating the unknown scaling factor by dividing all of them by any one of them that is non-zero such as the mean, maximum, root-mean-square, etc. This effectively yields $N^2-3N+1$ real-valued invariants after the loss of one degree of freedom in eliminating the unknown scale factor. For the EHT, if $N=7$ telescopes are operational, there are 29 closure invariants in an instantaneous snapshot.

\section{Morphological Classification}\label{sec:classification}

The first goal of this paper is to explore the possibilities of classifying image morphologies using simple ML methods entirely from interferometric closure invariants rather than visibilities because the latter are not immune to calibration errors. 

\begin{figure*}
\centering
\subfloat[][Point-like \label{fig:point-source}]
{\includegraphics[width=0.32\textwidth]
{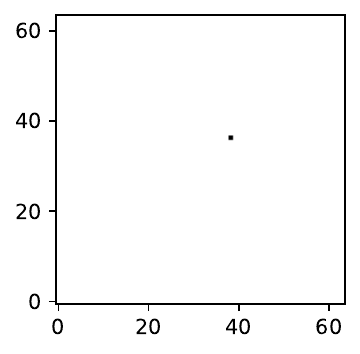}
}
\subfloat[][Uniform circular disc \label{fig:disc}]
{\includegraphics[width=0.32\textwidth]
{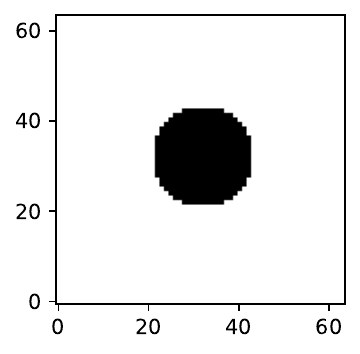}
}
\subfloat[][Double disc \label{fig:double-disc}]
{\includegraphics[width=0.32\textwidth]
{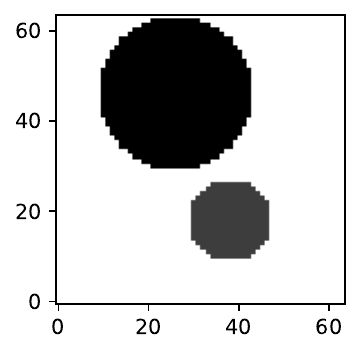}
} \\
\subfloat[][Crescent \label{fig:crescent}]
{\includegraphics[width=0.32\textwidth]
{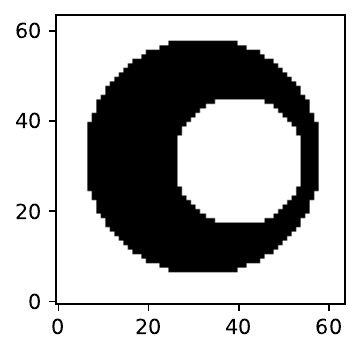}
}
\subfloat[][Crescent with accretion disc \label{fig:crescent-accretion}]
{\includegraphics[width=0.32\textwidth]
{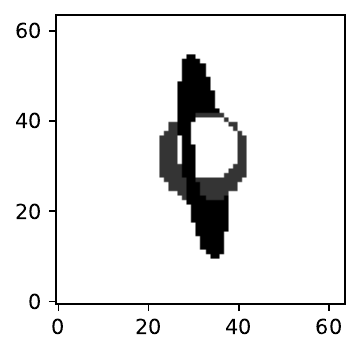}
}
\subfloat[][Crescent with jets \label{fig:crescent-jets}]
{\includegraphics[width=0.32\textwidth]
{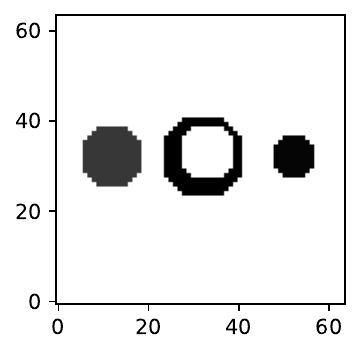}
}
\caption{Example images of six morphological classes that are classified using closure invariants. Each pixel corresponds to 3.52~$\mu$as. (a) Point-like objects are simulated at random locations with random intensities. (b) Uniform circular disc morphology is parametrised by its radius. (c) Double disc morphology is parametrised by the radii of the two discs, the $X$- and $y$-offsets between their centres, and the ratio of their intensities. (d) A crescent morphology is parametrised by the inner and outer radii, and the $X$- and $y$-offsets of the inner hollow region from the centre. (e) A crescent with accretion disc morphology is parametrised by the crescent parameters in (d), an elliptical accretion disc parametrised by semi-major axis with an ellipticity of 0.2 and a position angle, and an intensity ratio between the crescent and the accretion disc. (f) A crescent with double jet morphology consists of a crescent as parametrised above, in addition to two diametrically opposite jets with $x$- and $y$-offsets from the centre, and two intensities relative to the crescent structure. \label{fig:morphology-classes}}
\end{figure*}

We consider six different fiducial morphological classes (0--5) along with the parametrisations described below. Closure invariants are insensitive to some parameters such as absolute location and absolute intensity scale. The parameters that can be inferred for image reconstruction after the morphological classification are italicised: 
\begin{enumerate}[label=\arabic* --, start=0, leftmargin=2\parindent]
  \item Point-like: location, intensity~$\in[1,255]$,
  \item Uniform circular disc: \textit{radius} ($R\in[3, 32]$), constant intensity~$\in[1,255]$ throughout the disc,
  \item Dual circular discs: \textit{radii of the two discs} ($R_1\in[4,18]$, $R_2\in[3,11] < R_1$), \textit{offset of the centre of the second disc from the first} ($\delta x \in [-51,50]$, $\delta y \in [-22,21]$), intensity of first disc ($I_1$), \textit{ratio of intensities of the two discs} ($I_1/I_2 \in [1,300]$), 
  \item Crescent: \textit{radii of the outer and inner discs} ($R_1 \in [9,31]$, $R_2 \in [6,26] < R_1$), \textit{offset of the centre of the inner disc from the outer disc} ($\delta x \in [-22,18]$, $\delta y \in [-22, 18]$), intensity of outer disc ($I_1\in[1,255]$), intensity of inner disc is constrained to be $-I_1$,
  \item Crescent with elliptical accretion disc: crescent parameters as above ($R_1 \in [6,19]$, $R_2 \in [3,17] < R_1$), an elliptical accretion disc parametrised by \textit{semi-major axis} ($R_a\in[23,32]$), semi-minor axis fixed at $0.2 R_a$, a \textit{position angle} ($\theta_a\in[0,\pi]$), and an \textit{intensity ratio between the crescent and the accretion disc} ($I_\textrm{c}/I_\textrm{a} \in [0.5,1.5]$),
  \item Crescent with jet lobes: crescent parameters as above ($R_1 \in [6,13]$, $R_2 \in [3,12] < R_1$), two diametrically opposite jet lobes with \textit{radii} ($r_1\in[3,10], r_2\in[3,10]$) and \textit{offsets from the crescent centre} ($|\delta x_\textrm{jet}| \in [13,23]$, $|\delta y_\textrm{jet}| \in [13,23]$), and \textit{jet lobe intensities relative to the crescent intensity} ($I_1/I_\textrm{c} \in [0.006, 3]$, $I_2/I_\textrm{c} \in [0.006, 3]$).
\end{enumerate}
Note that these parametrisations are purely phenomenological and not physical. Fig.~\ref{fig:morphology-classes} shows examples of these morphological classes obtained with random instances of the respective parameters. The intensities are specified in arbitrary units, while linear dimensions are in pixels. The image sizes considered are $64\times 64$, with an angular resolution of 3.52~$\mu$as per pixel.

\subsection{Input measurements}\label{sec:inputs}

For sampling the fiducial Fourier (aperture) plane, we use frequencies of $\nu_1=230$~GHz and $\nu_2=345$~GHz. The \texttt{eht-imaging} software, which contains the exact locations of the EHT array elements \citep{eht19-2} provides our fiducial array layout. Two fiducial times of $t_1=\texttt{2017-04-05 04:46:05}$ UTC and $t_2=\texttt{2017-04-05 05:34:05}$ UTC are chosen to have seven locations of telescopes of the EHT array for which an object at RA=12$^\textrm{h}$30$^\textrm{m}$49.42$^\textrm{s}$, Dec=+12$^\circ$23$^\prime$28.04$^{\prime\prime}$ (J2000) would be visible simultaneously. The seven locations correspond to that of the Atacama Large Millimeter/submillimeter Array \citep[ALMA;][]{ALMA+2018} and the Atacama Pathfinder Experiment telescope \citep[APEX;][]{APEX+2015} in Chile, the Large Millimeter Telescope Alfonso Serrano \citep[LMT;][]{LMT+2016} in Mexico, the IRAM 30~m telescope on Pico Veleta \citep[PV;][]{PV+1995} in Spain, the Submillimeter Telescope Observatory \citep[SMT;][]{SMT+1999} in Arizona, the James Clerk Maxwell Telescope (JCMT) and the Submillimeter Array \citep[SMA;][]{SMA+2016} in Hawai'i. For this demonstrative study, we treat all telescopes as identical even though they exhibit substantial differences in their actual characteristics \citep{eht19-2}.

\begin{figure*}
\centering
\subfloat[][Sampled Fourier components in the aperture plane. \label{fig:uv}]
{\includegraphics[width=0.45\textwidth]
{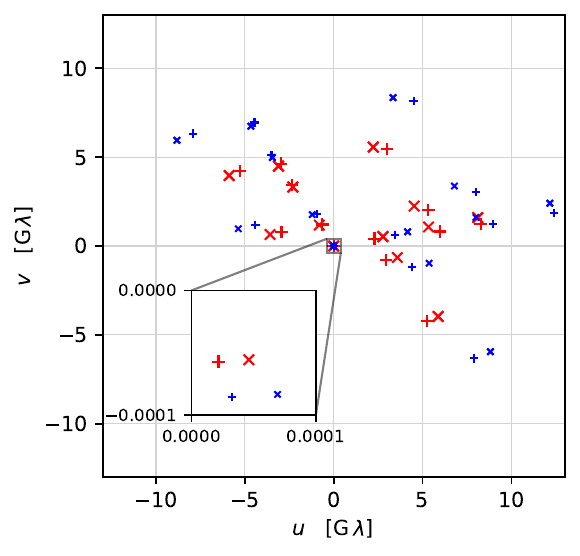}
}
\subfloat[][Closure invariants for various aperture models. \label{fig:closure-invariants}]
{\includegraphics[width=0.45\textwidth]
{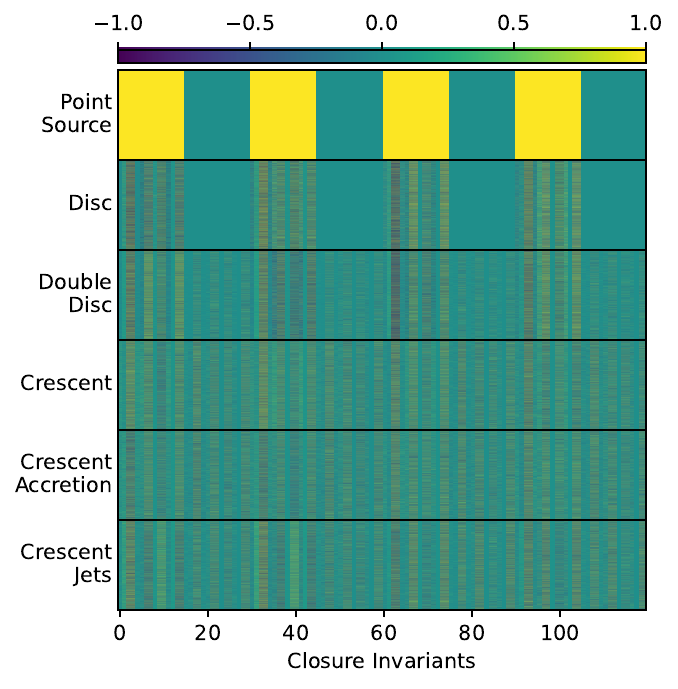}
}
\caption{(a) Aperture (Fourier) plane sampling (in units of wavelengths) obtained using locations of seven telescopes in the EHT array at $\nu_1=230$~GHz (red) and $\nu_2=345$~GHz (blue) at times $t_1=\texttt{2017-04-05 04:46:05}$ UTC ($+$ symbols) and $t_2=\texttt{2017-04-05 05:34:05}$ UTC ($\times$ symbols). The inset shows a zoomed view of the shortest element spacings. The three aperture sampling configurations in this paper correspond to $\nu_1$ at $t_1$ (red $+$ denoted by \texttt{1Fx1T}), $\nu_1$ at $t_1$ and $t_2$ (red $+$ and $\times$ denoted by \texttt{1Fx2T}), and $\nu_1$ and $\nu_2$ at $t_1$ and $t_2$ (red and blue $+$ and $\times$ denoted by \texttt{2Fx2T}). (b) Closure invariants for 10,000 random realisations of parameters under each of the six morphological classes as labelled on the $y$-axis. The closure invariants are indexed on the $x$-axis. The aperture models \texttt{1Fx1T}, \texttt{1Fx2T}, and \texttt{2Fx2T} correspond to the first 30, 60, and 120 indices, respectively. The values of closure invariants are indicated by the colour scale. Alternating bands of 15 indices correspond to the real and imaginary values of the complex advariants from which the real-valued co-polar closure invariants are derived using Abelian gauge theory formalism \citep{Thyagarajan+2022}, in which all point-like morphologies have zero imaginary values and unit real values regardless of their location or absolute intensities. \label{fig:input-measurements}}
\end{figure*}

Fig.~\ref{fig:uv} shows the aperture sampled in this study. The red and blue symbols denote 230~GHz and 345~GHz, respectively, while the $+$ and $\times$ symbols denote the two times at which the aperture was sampled, thereby yielding four possible combinations. We employ three sampling configurations in this paper to study the effect of varying the level of aperture plane sampling on the morphological classification accuracy. These aperture configurations correspond to time $t_1$ at frequency $\nu_1$, times $t_1$ and $t_2$ at frequency $\nu_1$, and times $t_1$ and $t_2$ at frequencies $\nu_1$ and $\nu_2$. They are denoted by \texttt{1Fx1T}, \texttt{1Fx2T}, and \texttt{2Fx2T}, respectively. The seven elements of the EHT array yield 21 snapshot visibilities at any given time and frequency. Therefore, the three sampling configurations consist of 21, 42, and 84 visibilities, respectively. We employ a sparser aperture sampling than in typical real observations to test the performance against sparse measurements.

We generate 10,000 instances of $64\times 64$ images for each morphological class using random realisations of their respective parameters (see Fig.~\ref{fig:morphology-classes}). The pixel resolution in angular units is 3.51~$\mu$as. Assuming all the array elements are identical and have uniform power sensitivity across the entire image, we simulated the visibilities using a direct Fourier Transform at the antenna spacings for the various realisations of the images under each of the morphological classes using equation~(\ref{eqn:rime}). At this point, we do not add noise to the simulated visibilities. We consider the impact of noise in \S\ref{sec:noise}.

The closure invariants are constructed from equation~(\ref{eqn:advariant}) using the Abelian gauge theory formalism for co-polar invariants \citep{Thyagarajan+2022}. In this formalism, a 7-element array yields 15 independent triads and a total of 30 real numbers (from 15 complex-valued \textit{advariants}) with one unknown but common scale factor shared between them. Dividing them all by any one of them (which is non-zero) eliminates this common scale factor and provides 29 closure invariants with the thirtieth number being the trivial unity as a result of this division as detailed in \citet{Thyagarajan+2022}. In this study, besides the 29 invariants, we also keep the additional trivial number -- the unity -- purely for bookkeeping convenience even though it contributes no valuable information. 

Fig.~\ref{fig:closure-invariants} shows the closure invariants computed for various realisations of each morphology. The classes are labelled on the left with horizontal lines denoting class boundaries, each of which includes 10,000 realisations. The closure invariants are numbered along the $x$-axis. Alternating bands of 15 closure invariants correspond to the real and imaginary parts of the advariants \citep{Thyagarajan+2022}, respectively. The three aperture sampling configurations, \texttt{1Fx1T}, \texttt{1Fx2T}, and \texttt{2Fx2T} correspond to the first 30, 60, and 120 closure invariants, respectively. The colour scale represents the value of the closure invariants. 

We note that for a point-like morphology, all the complex-valued advariants have real and equal values. Therefore, the closure invariants consist of equal numbers of ones and zeros. This is equivalent to having zero closure phases and unit closure amplitudes independent of the location or the absolute intensity, as expected from the \textit{centrosymmetry} of a point-like object \citep{TMS2017,Thyagarajan+2022}. Similarly, the symmetry of a uniform circular disc is reflected in the vanishing of all the imaginary parts. More complicated morphological classes have correspondingly non-trivial combinations of real and imaginary parts. 

The input for the training of our ML classifiers and parameter estimators comprises of these closure invariants, the morphological class labels and class parameters they were generated from. The ML model so generated then takes an unknown set of closure invariants as input and predicts the morphological class and the respective parameters. Note that neither the images nor the visibilities themselves are used as inputs.

\subsection{Machine Learning Models} \label{sec:ML-classifiers}

We consider four simple ML methods for classifying the image morphologies, whose details and performance are characterised below. We use \texttt{PyTorch} for implementing most of the models except for the Random Forest classifier for which we use \texttt{scikit-learn}. The ML models are visually illustrated in Fig.~\ref{fig:classification-graphs} in Appendix~\ref{sec:ML-graphs}. In each of these methods, we split the data randomly with 80\% used for training and 20\% for validation. We randomise the composition of the training and validation sets several times maintaining the same proportional split, and find no significant change in the results. For each method, we choose the minimum possible number of training epochs soon after the training loss flattens to an extent of showing no significant change ($\lesssim 0.1$\%) to avoid over-fitting to the training data and to preserve the method's generalisation capabilities for unseen test data. 

\subsubsection{Logistic Regression} \label{sec:LR}

Logistic regression (LR) estimates the probability of an event occurring, such as the morphological class, based on a given data set of independent variables, namely, the closure invariants. LR is considered a discriminative model, which means that it attempts to distinguish between classes. Since the dependent variable (morphological class) has six possible outcomes, and no specified order, we employ multinomial LR, which is a simple network consisting two layers -- input (corresponding to the number of closure invariants) and output (number of morphological classes).

\begin{figure*}
\centering
\subfloat[][Logistic Regression \label{fig:cmat-LR}]
{\includegraphics[width=0.33\textwidth]
{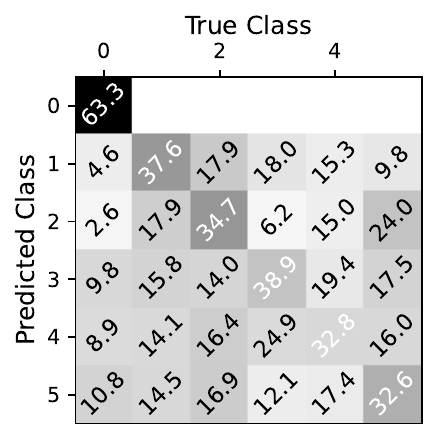}
}
\subfloat[][Multi-layer Perceptron \label{fig:cmat-MLP}]
{\includegraphics[width=0.33\textwidth]
{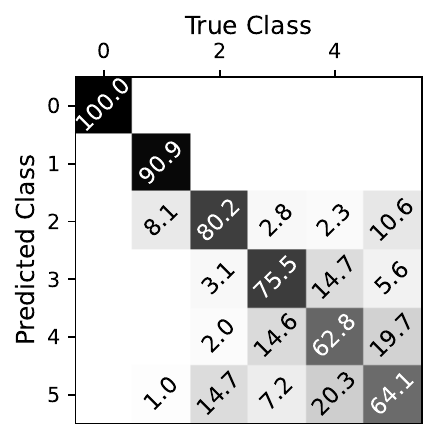}
}
\subfloat[][Random Forest \label{fig:cmat-RF}]
{\includegraphics[width=0.33\textwidth]
{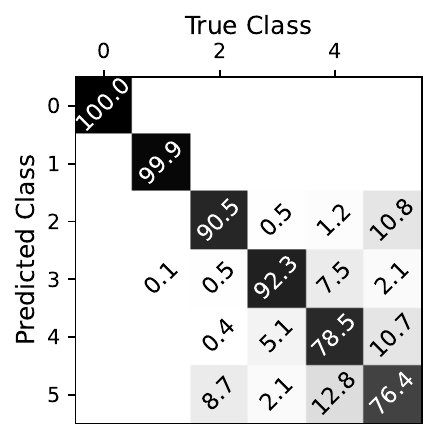}
} 
\caption{Confusion matrices (in percentage) of \textit{(a)} Logistic Regression, \textit{(b)} Multi-layer Perceptron, and \textit{(c)} Random Forest from training for image morphology classification. Diagonals denote accurate classifications while off-diagonals denote misclassifications. \label{fig:cmats}}
\end{figure*}

The training is performed using \textit{Sigmoid} activation, \textit{Adam} optimiser (learning rate $10^{-4}$), and 10 epochs. Fig.~\ref{fig:cmat-LR} shows the confusion matrix\footnote{In statistical classification, a confusion matrix is a $M\times M$ matrix consisting of statistics of ground truths and predictions for $M$ classes. The diagonal elements represent statistics of accurate classification while off-diagonals denote that of misclassifications.} between true and predicted classes using LR for the \texttt{2Fx2T} aperture model. The presence of significant off-diagonal values indicates that LR's classification was not very accurate even though it is notably better than a random classification. Even point-like morphologies are accurately classified in only 63\% of the instances. The classification performance on other morphologies is only $\sim 33-38$\%. LR is one of the simplest classification schemes and is not expected to perform with high accuracy on non-binary classes. 

\subsubsection{Multi-layer Perceptron} \label{sec:MLP}

Multi-layer Perceptron (MLP), a supervised learning algorithm, is a fully connected feed-forward artificial neural network with at least three layers (input, output, and at least one hidden layer) that can learn to classify non-linearly separable patterns. 

We perform a 15-epoch training with an \textit{Adam} optimiser (learning rate $10^{-3}$) and a Rectified Linear Unit (\texttt{ReLU}) non-linear activation in each of the hidden layers. Fig.~\ref{fig:cmat-MLP} shows the confusion matrix obtained for the \texttt{2Fx2T} aperture model. Relative to LR's performance, the confusion matrix is much tighter around the diagonal. Point-like morphologies are perfectly predicted. The uniform circular disc morphology is predicted accurately at the rate of 91\%. The prediction accuracy decreases with increasing complexity of the underlying morphology. The worst performance is seen in the case of morphology classes 4 (crescent with accretion disc) and 5 (crescent with jet lobes) at a prediction accuracy of $\simeq 63-64$\%, which is still better than the best performance of LR in the case of point-like morphology. 

\subsubsection{Random Forest} \label{sec:RF}

A random forest classifier (RF) is an estimator that fits a number of decision tree classifiers on various sub-samples of the data set and uses averaging to improve the predictive accuracy and control over-fitting. We use 200 trees in the forest. Fig.~\ref{fig:cmat-RF} shows the confusion matrix obtained with training on the \texttt{2Fx2T} aperture model with an RF classifier. The point-like and uniform circular disc morphologies are classified perfectly. The double disc and crescent morphologies are classified at $\gtrsim 90$\% accuracy, and the crescent with accretion disc and jet lobes are classified with $\gtrsim 76$\% accuracy. Overall, for the specific sets of trainable parameters chosen in this work, the RF classifier appears to perform better than the rest. 

\subsection{Effect of aperture coverage on classifiers' performance}\label{sec:aperture-coverage}

We train the aforementioned models for the three aperture models with different levels of aperture coverage with a balanced input of morphological classes. To test their performance on unknown test data, we use the $F_1$ score \citep{F-score}, a harmonic mean of \textit{precision} and \textit{recall}, as a measure of the test's accuracy. During testing, to avoid skewing the results based on the level of balance of input data, we use 1000 iterations each containing 100 randomly drawn test inputs (excluding the training set) with no regard for the balance of classes to determine the $F_1$ score statistics even including the worst cases of imbalance across classes.  

\begin{figure}
\centering
\includegraphics[width=\linewidth]
{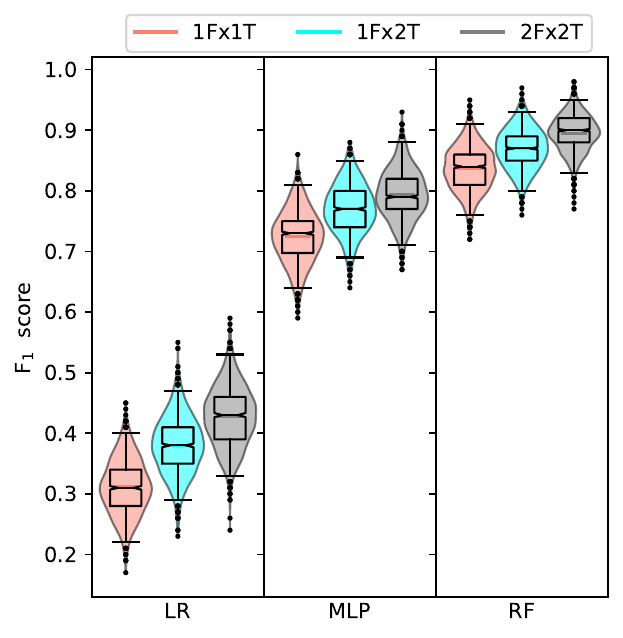}
\caption{$F_1$ score distributions for our ML classification models -- Logistic Regression (LR), Multi-layer Perceptron (MLP), Random Forest (RF) -- from a randomised balance of input morphologies for aperture coverage models \texttt{1Fx1T} (orange, left violin plot), \texttt{1Fx2T} (cyan, middle violin plot), and \texttt{2Fx2T} (grey, right violin plot). The box-whisker plots show median, lower and upper quartiles, 95\% confidence limits, and outliers.}
\label{fig:F1-aperture}
\end{figure}

Fig.~\ref{fig:F1-aperture} shows the distribution of $F_1$ scores for randomly drawn test inputs. LR and RF appear to have the worst and best $F_1$ scores, respectively. MLP has intermediate $F_1$ scores. In all cases, the $F_1$ scores show marked improvement as the aperture coverage improves. This is because the morphological information sampled by the successively improving aperture coverage helps build better predictive ML classifiers. 

Even with limited aperture coverage (Fig.~\ref{fig:uv}), we find that the accuracy in a multi-class classification can be $>80$\% using simple ML models. Increasing the aperture coverage is expected to improve the classification accuracy further. Refinements to these models for improved classification accuracy is left to future work. 

\subsection{Effect of noise} \label{sec:noise}

We examine the effect of noise in the data by injecting various levels of Gaussian noise into the images. We quantify the noise through the inverse of signal-to-noise ratio (SNR) at levels $\textrm{SNR}^{-1}=0$ (noiseless), $1/30$, $1/10$, and $1/3$, with the noise standard deviation defined relative to the maximum intensity in the images. The images are subsequently converted to visibilities, and eventually to closure invariants that are used for training and validation. Noisy test closure invariants data are then passed to the trained models to determine the classification accuracy, $F_1$. Note that this noise in the closure invariants is non-Gaussian due to the non-linear transformations.

\begin{figure}
\centering
\includegraphics[width=\linewidth]
{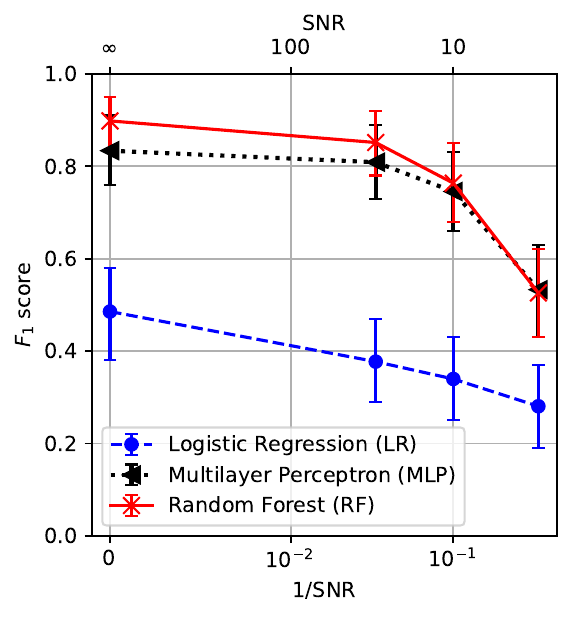}
\caption{$F_1$ score variation as a function of injected noise in test inputs, denoted by $\textrm{SNR}^{-1}$, for logistic regression (blue dashed line connecting filled circles), multi-layer perceptron (black dotted line connecting filled triangles), and random forest (red solid line connecting crosses) models that were trained on noisy data using the \texttt{2Fx2T} aperture coverage model. The error bars denote 95\% confidence intervals. The $x$-axis uses a linear-logarithmic scaling. The $F_1$ scores of our ML models trained on noiseless data but tested on noisy inputs are presented in Fig.~\ref{fig:F1-noiseless-on-noise}.}
\label{fig:F1-noise}
\end{figure}

Fig.~\ref{fig:F1-noise} shows the impact of injected noise levels on $F_1$ scores. Overall, the MLP and RF models perform similarly with the latter achieving marginally better accuracy. The LR model performs at roughly half the accuracy as the other two. The performance of the MLP and RF models at an $\textrm{SNR}=30$ are comparable in the noiseless case with accuracies of $\sim$ 80--90\%, which drop to $\sim 50\%$ when $\textrm{SNR}=3$.

It is seen that the MLP model trained on noisy and noiseless (infinite SNR) data and applied on noiseless test data ($\textrm{SNR}^{-1}=0$ in Fig.~\ref{fig:F1-noise}) performs slightly better (by a few percent) than that trained only on noiseless data (\texttt{2Fx2T} model in Fig.~\ref{fig:F1-aperture}) due to the additional diversity of information provided through the varying SNR in the training and validation process. A worst-case scenario of the ML models' performance when trained on noiseless data but tested on noisy data is shown in Fig.~\ref{fig:F1-noiseless-on-noise}.

The EHT has non-uniform noise levels across its baselines because the SEFD of the participating telescopes vary significantly \citep{eht19-2}. In this work, as noise is injected through the images, the noise levels in the visibilities are approximately uniform across element separations. However, the wide range of SNR covered and the corresponding accuracy range recorded here can be indicative of the accuracy achievable for non-uniform noise levels such as in the EHT. More complex scenarios like baseline-dependent noise levels applicable to specific cases like the EHT are being explored separately.

\subsection{Performance on untrained class data} \label{sec:untrained-class}

We conduct a limited study of applying our trained models on test data drawn from two untrained classes, namely, $m$-ring models \citep{Johnson+2020,Roelofs+2023}, and radiatively inefficient accretion flow (RIAF) models of Sgr~A$^\star$ \citep{Broderick+2011}.

\begin{figure}
\centering
\subfloat[][Classification of $m$-rings \label{fig:mring-classification}]
{\includegraphics[width=\linewidth]
{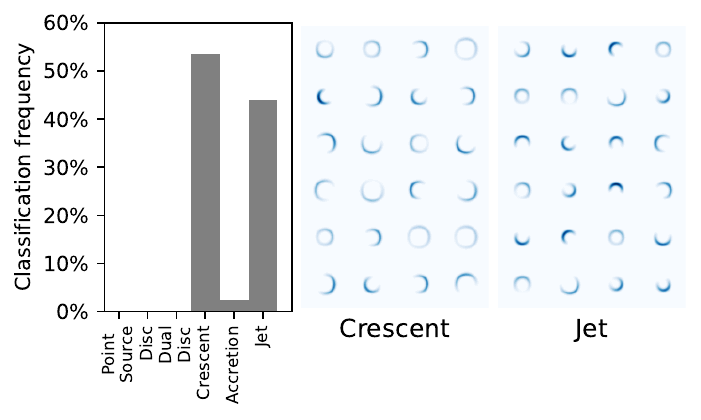}
} \\
\subfloat[][Classification of Sgr~A$^\star$ images from RIAF models \label{fig:sgra-classification}]
{\includegraphics[width=\linewidth]
{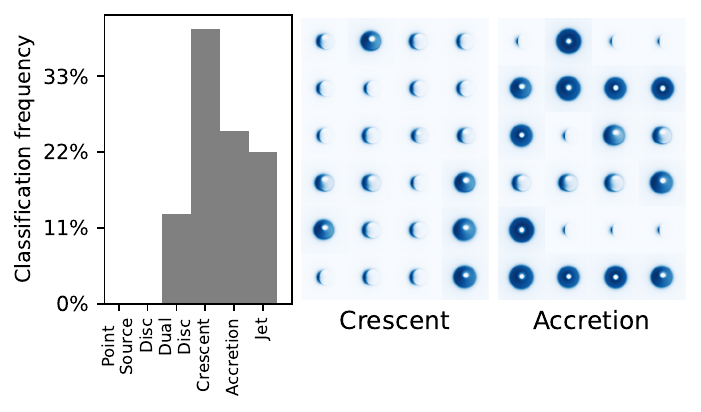}
}
\caption{(a) MLP model classification of $m$-ring data (untrained class). The $m$-ring test data (order $m=1$) are predominantly classified as belonging to the crescent and jet classes. A few examples of $m$-ring data classified as crescents and jets are shown in the middle and right panels, respectively. (b) MLP model classification of Sgr~A$^\star$ images from radiatively inefficient accretion flow (RIAF) models (another untrained class). The RIAF Sgr~A$^\star$ test data are predominantly classified as belonging to the crescent, accretion, and jet classes. A few examples of the RIAF data classified under crescent and accretion categories are shown in the middle and right panels, respectively. Note that our jet and accretion classes include a crescent as well (see Fig.~\ref{fig:morphology-classes}).}
\end{figure}

We explored $m$-ring models of various orders, but present results here only for order $m=1$ with the $\beta$ parameter, $\beta_k=0$, for $|k|>1$. Fig.~\ref{fig:mring-classification} illustrates the classification performance of our MLP model for first order $m$-ring models drawn from 1000 randomly sampled parameters. They are classified primarily as crescents and jets in $\approx 54$\% and $\approx 43$\% of the instances, respectively. A few examples of the input test images classified as crescents and jets are also shown.


It is noteworthy that the $m$-ring data classified as jets clearly have no lobe-like features as in Fig.~\ref{fig:crescent-jets}. We attribute this confusion in classification to the fact that our jet class definition includes a crescent in addition to the jet lobes, which will become indistinguishable from the crescent class when the jet lobes are significantly faint. Therefore, the two are not orthogonal but overlapping classes, which are known to suffer from ambiguity.

Notably, the images classified as jets have systematically smaller diameters than those classified as crescents. This is because the jet lobes in the training have to fit inside the image's fixed bounding box and thus restricts the crescents to smaller sizes (see parameter ranges in \S\ref{sec:classification}). Thus, it is apparent that our classification model assigns the crescent's dimensions a greater degree of importance over the presence of jet lobes in the data.

Fig.~\ref{fig:sgra-classification} shows the MLP's classification performance on 9090 instances of test data drawn from the RIAF Sgr~A$^\star$ images. The majority ($\approx 40$\%) are classified as crescents, but a significant number of cases ($\approx 22$\%) are also classified as accretion and jets. Notably, these classes also include a crescent.

In either case, the confusion in classification arises primarily due to the overlap between morphological classes. The confusion between these classes is also evident from the confusion matrices in Fig.~\ref{fig:cmats}.

Tackling overlapping classes requires more careful consideration of class definitions, and other advanced techniques like feature engineering that involves enhancing emphasis on discriminative features, ensemble learning that trains different classification algorithms and combines their predictions, and active learning requiring the user to provide inputs on ambiguous data points. These avenues will be explored in future work.

\section{Morphological Parametrisation} \label{sec:parametrisation}

Here, we investigate the ability of ML regression to estimate parameters that are retrievable using only closure invariants and implement a parametric image reconstruction. Absolute position and intensity information is lost while using closure invariants, and we don't attempt to retrieve such information. We approach this parametrisation not in sequence to the classification but as an independent subject. The morphological class is assumed to have been specified independently. Our approach is to parametrically reconstruct an image once that classification is specified by independent means. Here, we consider the \texttt{2Fx2T} aperture model. 

Point-like morphologies can only be detected and classified, but cannot be reconstructed because they are parametrised by a single location and single intensity, whose absolute values are not accessible by closure invariants. We study the parametric image reconstruction of the following morphological classes: uniform circular disc (1), dual disc (2), and crescent (3), noting that we do not estimate an absolute location or an absolute intensity scale.  We employ MLP to determine the morphological parameters. The MLP models used for the different morphologies are visually illustrated Fig.~\ref{fig:parametrisation-graphs} in Appendix~\ref{sec:ML-graphs}. The specific MLP model is briefly described below for each morphology. 

\subsection{Uniform circular disc}\label{sec:circular-disc}

\begin{figure}
\centering
\includegraphics[width=\linewidth]
{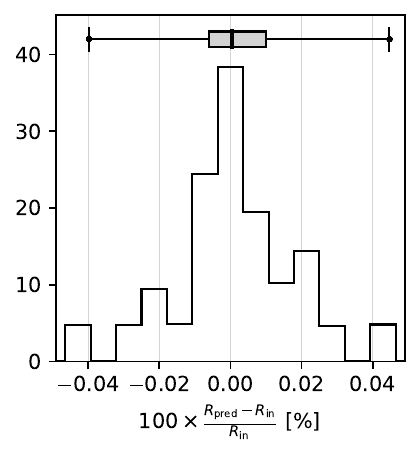}
\caption{Histogram (in percent) of percentage error in the prediction of uniform circular disc radius using MLP (see Fig.~\ref{fig:disc-graph}). The box-whisker plot shows the median, lower and upper quartiles, and the 95\% confidence interval ($\simeq \pm 0.04$\%) of the prediction error.}
\label{fig:Disc-R}
\end{figure}

The only parameter of the uniform circular disc that we can estimate with closure invariants is its radius. We use an an MLP model with \texttt{ReLU} activation and an \textit{Adam} optimiser (learning rate $10^{-3}$). After 10 epochs, the prediction accuracy asymptotically approaches 100\%. Fig.~\ref{fig:Disc-R} shows a histogram of the relative error between the prediction and the true value in percent. The prediction is accurate to within 0.04\% with 95\% confidence. 

\subsection{Crescent}\label{sec:crescent}

In the morphologies we studied, the crescent model offers the next level of complexity relative to the uniform circular disc. The parameters of the crescent model we estimate are the radii of the outer and inner discs, $R_1$ and $R_2$, respectively, and the $x$- and $y$-offsets of the centre of the inner disc from the outer one, $\delta x$, and $\delta y$, respectively. We use an MLP model with \texttt{ReLU} activation, 20\% dropout of neurons between layers, and an \textit{Adam} optimiser (learning rate $10^{-3}$). We perform a 10-epoch training with 8-fold cross-validation.

\begin{figure*}
\centering
\subfloat[][Outer disc radius, $R_1$ \label{fig:cr-R1}]
{\includegraphics[width=0.32\textwidth]
{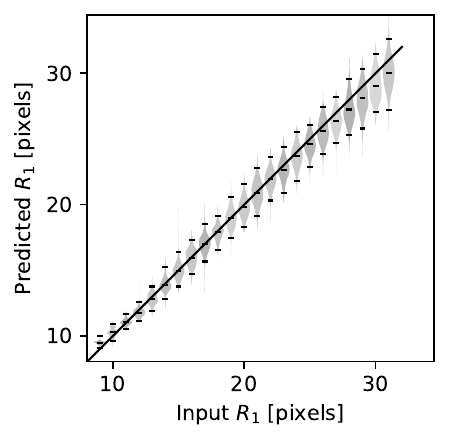}
}
\subfloat[][Inner disc radius, $R_2$ \label{fig:cr-R2}]
{\includegraphics[width=0.32\textwidth]
{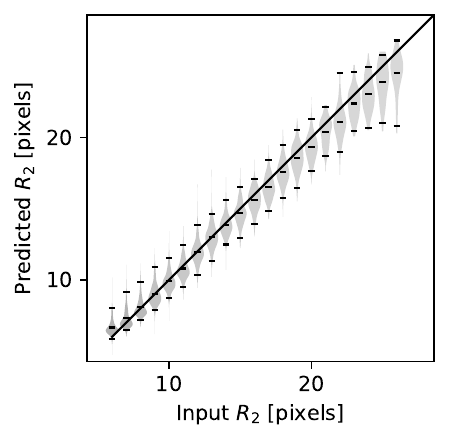}
} \\
\subfloat[][Relative $x$-offset between two discs, $\delta x$ \label{fig:cr-dX}]
{\includegraphics[width=0.32\textwidth]
{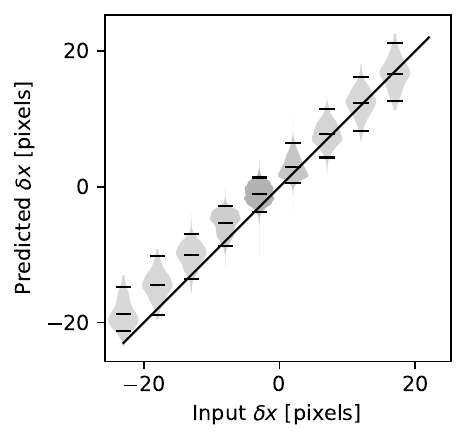}
}
\subfloat[][Relative $y$-offset between two discs, $\delta y$ \label{fig:cr-dY}]
{\includegraphics[width=0.32\textwidth]
{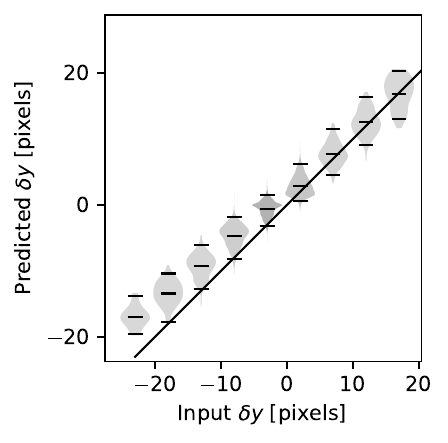}
} \\
\subfloat[][Input image example \label{fig:cr-inp}]
{\includegraphics[width=0.32\textwidth]
{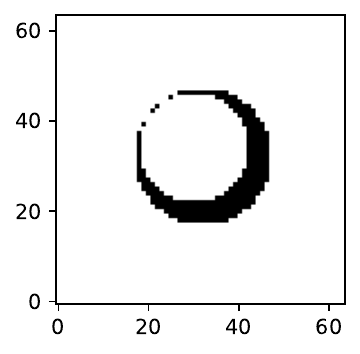}
}
\subfloat[][Parametrically reconstructed image\label{fig:cr-pred}]
{\includegraphics[width=0.32\textwidth]
{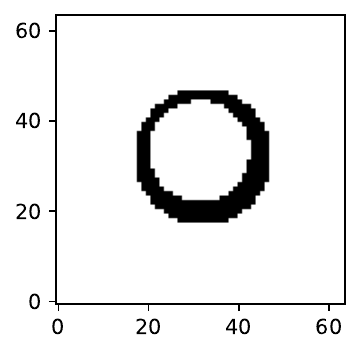}
}
\subfloat[][Difference image\label{fig:cr-diff}]
{\includegraphics[width=0.32\textwidth]
{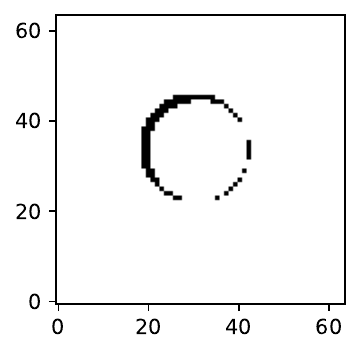}
}
\caption{Estimated parameters of crescent using MLP (see Fig.~\ref{fig:crescent-graph}) and parametric image reconstruction example. Predicted parameters against input parameters are shown for \textit{(a)} outer disc radius, $R_1$, \textit{(b} inner disc radius, $R_2$, \textit{(c)} $x$-offset, $\delta x$, and \textit{(d)} $y$-offset, $\delta y$ between the two discs. The diagonal lines denote perfect prediction. The mean and central 95th percentile are marked on the ``violin'' plots. The shade of the ``violin'' indicates the relative number of input data in that bin. \textit{(e)} Input crescent image example. \textit{(f)} Parametrically reconstructed crescent image from estimated parameters. \textit{(g)} Difference between input and reconstructed image. \label{fig:crescent-prediction}}
\end{figure*}

Figs.~\ref{fig:cr-R1}--\ref{fig:cr-dY} show the predicted parameters against the true values as ``violin'' plots. The predictions of all the four parameters are consistent with their respective input values to within 95\% confidence over the range of values considered.  

As an example, Fig.~\ref{fig:cr-inp} shows an input image of a crescent. With the predicted parameters, Fig.~\ref{fig:cr-pred} shows a parametrically reconstructed image. The difference between the input and the parametrically reconstructed image is shown in Fig.~\ref{fig:cr-diff}. While the reconstructed image resembles the input image reasonably, the prediction accuracy is not as high as the uniform circular disc case. This is because the same amount of input data is used to constrain a larger number of parameters, thereby leading to a slight degradation in prediction accuracy. 

\subsection{Dual disc}\label{sec:dualdisc}

A double disc is more complex than a crescent because in addition to the radii of the two discs ($R_1$ and $R_2$) and the displacement of the centre of one relative to the other ($\delta x$ and $\delta y$), the ratio of two intensities ($I_1/I_2$) is an extra parameter. Again, we use an MLP model with \texttt{ReLU} activation, 20\% dropout of neurons between layers, and an \textit{Adam} optimiser (learning rate $10^{-3}$). We perform a 10-epoch training with 8-fold cross-validation.

\begin{figure*}
\centering
\subfloat[][Larger disc radius, $R_1$ \label{fig:dd-R1}]
{\includegraphics[width=0.32\textwidth]
{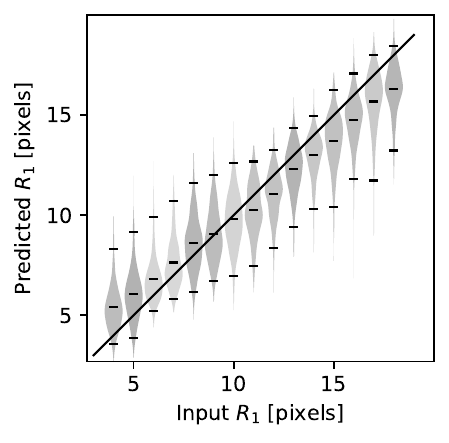}
}
\subfloat[][Smaller disc radius, $R_2$\label{fig:dd-R2}]
{\includegraphics[width=0.32\textwidth]
{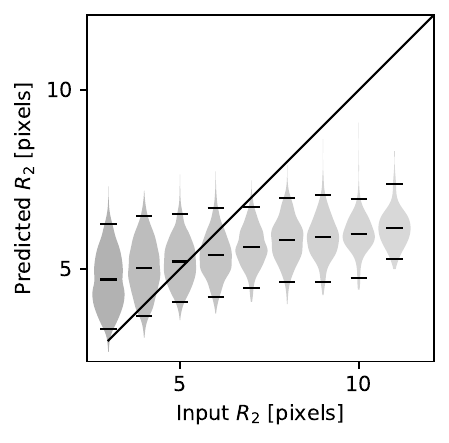}
}
\subfloat[][Intensity ratio, $I_1/I_2$ \label{fig:dd-Iratio}]
{\includegraphics[width=0.32\textwidth]
{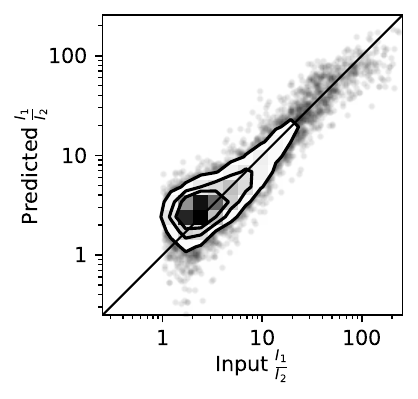}
} \\
\subfloat[][Relative $x$-offset between two discs, $\delta x$ \label{fig:dd-dX}]
{\includegraphics[width=0.32\textwidth]
{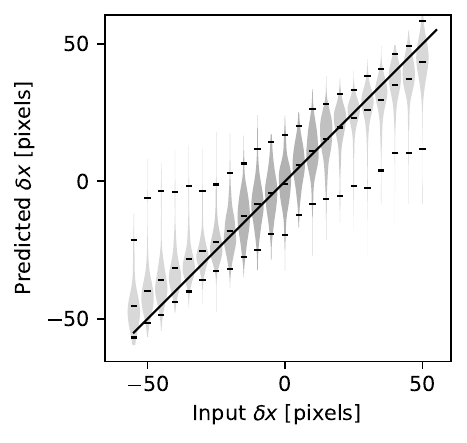}
}
\subfloat[][Relative $y$-offset between two discs, $\delta y$ \label{fig:dd-dY}]
{\includegraphics[width=0.32\textwidth]
{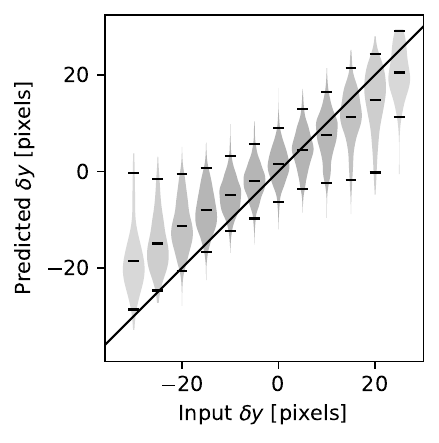}
} \\
\subfloat[][Input image example\label{fig:dd-inp}]
{\includegraphics[width=0.32\textwidth]
{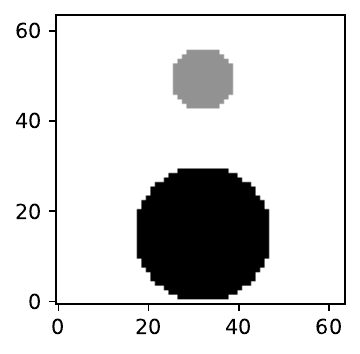}
}
\subfloat[][Parametrically reconstructed image\label{fig:dd-pred}]
{\includegraphics[width=0.32\textwidth]
{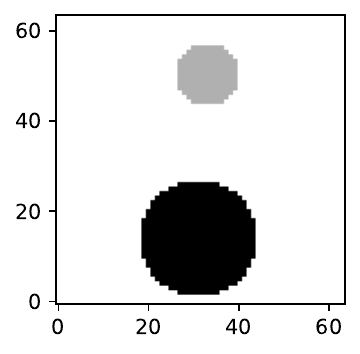}
}
\subfloat[][Difference image\label{fig:dd-diff}]
{\includegraphics[width=0.32\textwidth]
{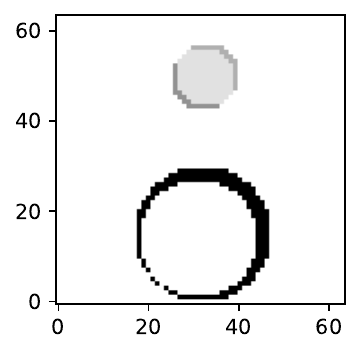}
}
\caption{Estimated parameters of dual disc using MLP (see Fig.~\ref{fig:dualdisc-graph}) and parametric image reconstruction example. Predicted parameters against input parameters are shown for \textit{(a)} larger disc radius, $R_1$, \textit{(b)} inner disc radius, $R_2$, \textit{(c)} ratio of two disc intensities, $I_1/I_2$, \textit{(d)} $x$-offset, $\delta x$, and \textit{(e)} $y$-offset, $\delta y$ between the two discs. The diagonal lines and ``violin'' plots have same meaning as in Fig.~\ref{fig:crescent-prediction}. Panels \textit{(f)}, \textit{(g)}, and \textit{(h)} show input, reconstructed, and difference images, respectively, of a dual disc example. \label{fig:dualdisc-prediction}}
\end{figure*}

Figs.~\ref{fig:dd-R1}--\ref{fig:dd-dY} show the input against the predicted parameters as violin plots. The ideal prediction is enveloped by the central 95\% of the distribution of predicted parameters in all cases except $R_2$. The model suffers from an over-prediction of $R_2$ at smaller values and a significant under-prediction at larger values. The prediction accuracy appears to have worsened relative to the crescent case. This is again attributed to an increase in the number of morphological parameters for the same number of input closure invariants, which results in inaccurate constraints on $R_2$. Figs.~\ref{fig:dd-inp}--\ref{fig:dd-diff} show an input, the parametric reconstruction, and the difference image for the dual disc model.  

To understand the origin of this systematic misestimation, we examine the \textit{Jacobian} matrix, $\mathbf{J}\coloneqq \partial C_i/\partial p_j$ that describes the sensitivity of closure invariants, $C_i$, to changes in morphological parameters, $p_j$, to explore degeneracies in the system. In the case of the \texttt{2Fx2T} aperture and dual disk models, we have 120 closure invariants (only 116 are independent and the rest kept for bookkeeping convenience) and 5 dual disc morphology parameters. Thus, $\mathbf{J}$ is a $120\times 5$ matrix. The rank of $\mathbf{J}$ indicates the degree of independence of the parameters in determining the closure invariants. A reduction from a full rank (5 in this case) corresponds to a commensurate number of degeneracies between the parameters, and a full rank implies that the closure invariants are sensitive to all 5 parameters with no degeneracies between them. We construct the \textit{Jacobian}, $J$, by introducing small perturbations to the chosen parameters and recording the corresponding changes in the closure invariants. We compute the rank of $\mathbf{J}$ by estimating the number of non-negligible singular values through Singular Value Decomposition (SVD). We repeat the rank determination over 1000 random realisations of the morphological parameters sampling various regions of the parameter space.

\begin{figure*}
\centering
\includegraphics[width=0.8\linewidth]
{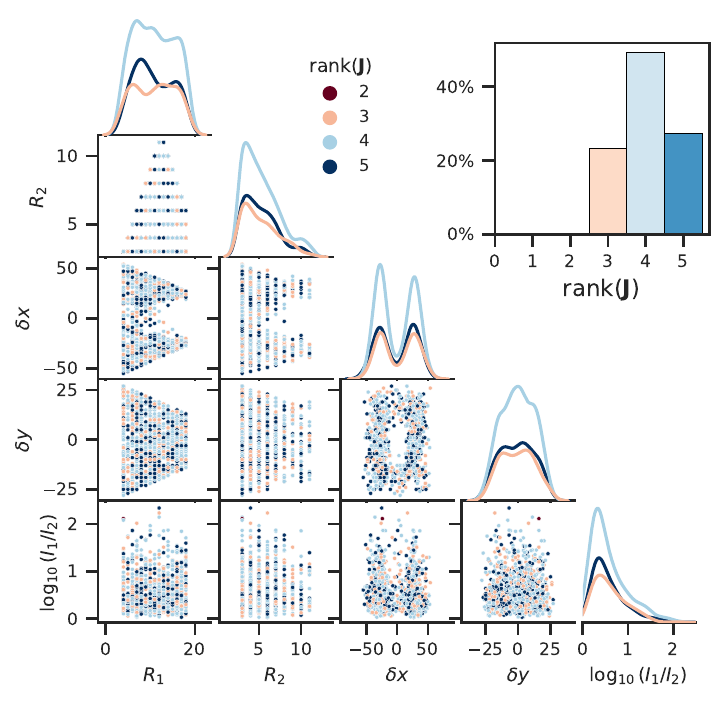}
\caption{Distribution of ranks of the \textit{Jacobian} matrix, $\mathbf{J}\coloneqq \partial C_i/\partial p_j$, between closure invariants, $C_i$, for the \texttt{2Fx2T} aperture model and 1000 randomly sampled parameters, $p_j$, for a dual disc morphology. The top right panel shows the overall distribution of the \textit{Jacobian's} ranks. A majority ($\approx 75$\%) are short of a full rank of 5 indicating significant degree of degeneracy between the morphological parameters in determining closure invariants. The \textit{corner plot} shows the distribution of ranks as a function of individual input dual disc parameters sampled.}
\label{fig:DualDisc-JacobianRank}
\end{figure*}

Fig.~\ref{fig:DualDisc-JacobianRank} shows a distribution of the \textit{Jacobian's} ranks (top right) from the 1000 realisations sampling the dual disc morphology. In a majority of instances ($\approx 75$\%), the \textit{Jacobian}, $\mathbf{J}$, is found to be short of full rank indicating a significant degree of degeneracy, where certain combinations of parameters are inconsequential in affecting closure invariants. The \textit{corner plot} illustrates the ranks as a function of sampled parameters shown pairwise with the marginal distributions shown on the diagonal. We are unable to identify clear trends of the behaviour of \textit{Jacobian} rank as a function of input parameters. This confirms that complex degeneracies are present in the data considered here. Another plausible origin of the degeneracy is the rather sparse aperture coverage used here. Using more data by enhancing the aperture beyond the sparse coverage considered here could potentially resolve these degeneracies. Refinement of ML models in conjunction with use of larger data size for mitigating the degeneracies and improving the prediction accuracy will be subject of future work.

\section{Discussion and Summary}\label{sec:summary}

Extreme VLBI conditions such as the requirement of high-precision calibration, sparse data, and low signal-to-noise pose a big challenge to interferometric imaging. In such cases, closure invariants, which are immune to calibration errors, critically provide a reliable information anchor for estimating the image morphology and preventing divergent solutions and artefacts, and have indeed been successfully employed for decades. Despite their extensive use, a direct backward transformation from closure invariants to morphological parameters has not been straightforward. This, and the prospect that closure invariants can provide independent avenues for constraining image morphology, are the primary motivators of this study. 

We demonstrate that Machine Learning methods applied on the closure invariants can be used to classify the image morphology as well as independently determine the morphological parameters if the morphological class is known \textit{a priori}. The latter can aid in parametric image reconstruction. We study the ability of simple methods like logistic regression, multi-layer perceptron, and random forest classifier to classify six different image morphologies -- point-like, uniform circular disc, dual disc, crescent, crescent with elliptical accretion disc, and crescent with double jet lobes. Among them, all except logistic regression yield $F_1$ scores $\gtrsim 0.8$ even with relatively sparse aperture coverage. The classification accuracy  of all ML models notably improves with increasing aperture coverage {and degrades with increasing noise. Our classifiers exhibit ambiguity between the crescent, jet, and accretion classes on untrained test inputs from $m$-ring and Sgr~A$^\star$ models. We attribute this to the fact that the class definitions used herein are not orthogonal but overlapping.

As a separate study, we employ simple multi-layer perceptron models to estimate morphological parameters such as radii, relative offsets, ratio of intensities, etc. for the uniform disc, crescent, and dual disc morphologies. With these estimates, we perform parametric image reconstruction without information about absolute position or intensity scale. The estimated parameters and reconstructed images are largely consistent and significantly correlated with the corresponding true values. Increasing the morphological complexity while using a relatively sparse aperture coverage results in a corresponding loss of accuracy in the estimated parameters. This is evidently due to the insensitivity of closure invariants to certain combinations of parameters owing to inherent degeneracies between them. Increasing the aperture coverage and the sophistication of the ML models is expected to resolve these degeneracies and improve the prediction accuracy, which will be subject of future study. 

Our proof-of-concept method demonstrates, using a backward-mapping approach, that both constraints on object morphologies and reconstruction of images are possible using interferometric closure invariants. It offers an independent approach that may be particularly suited for sparsely-sampled observations with challenging calibration requirements such as the EHT.  Machine learning methods can be useful in mitigating the degeneracies imposed by local minima and sparsity of information that may be encountered in existing methods when mapping from closure invariants to image morphologies. The potential of this method demonstrated herein warrants further exploration, the next steps of which will be to characterise its performance relative to other imaging methods. This approach could be used to complement, rather than replace, traditional methods since closure invariants do not carry any inherent information about absolute positions or intensities, and potentially other information due to the lost degrees of freedom relative to having reliably calibrated visibilities. 

\section*{Acknowledgements}

Inputs from John Carilli, Foivos Diakogiannis, Ron Ekers, Timothy Galvin, Aidan Hotan, Samuel Lai, Daniel Mitchell, John Morgan, Bojan Nikolic, Rajaram Nityananda, Cheng Soon Ong, and Maxim Voronkov are gratefully acknowledged.

\textit{Software:} \texttt{eht-imaging} \citep{eht-imaging}, \texttt{joblib} \citep{joblib}, \texttt{Matplotlib} \citep{matplotlib}, \texttt{NumPy} \citep{numpy}, \texttt{PyTorch} \citep{PyTorch2019}, \texttt{Scikit-learn} \citep{scikit-learn}, \texttt{SciPy} \citep{scipy}, \texttt{Seaborn} \citep{seaborn}, \texttt{Torchview} \citep{torchview}.

\section*{Data Availability}

The data underlying this article will be shared on reasonable request
to the corresponding author.




\bibliographystyle{rasti}


\appendix

\section{Visualisation of ML models} \label{sec:ML-graphs}

ML models used in the classification and parametrisation processes are illustrated graphically in Fig.~\ref{fig:classification-graphs} and \ref{fig:parametrisation-graphs}, respectively. 

\begin{figure*}
\centering
\subfloat[][Logistic Regression\label{fig:lr-graph}]
{\includegraphics[width=0.4\textwidth]
{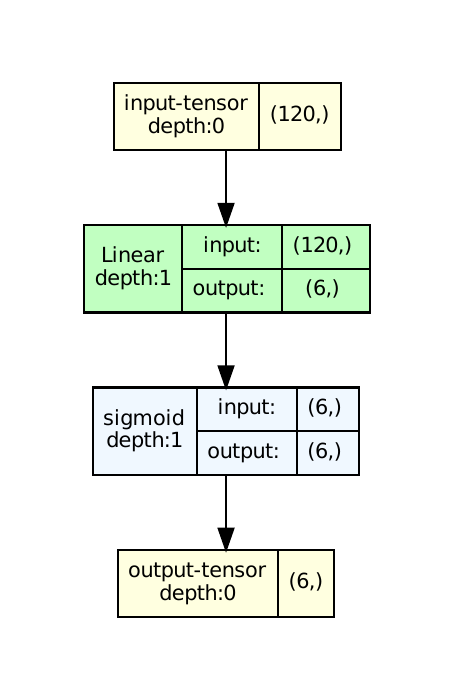}
}
\subfloat[][Multilayer Perceptron\label{fig:mlp-graph}]
{\includegraphics[width=0.35\textwidth,height=0.95\textheight]
{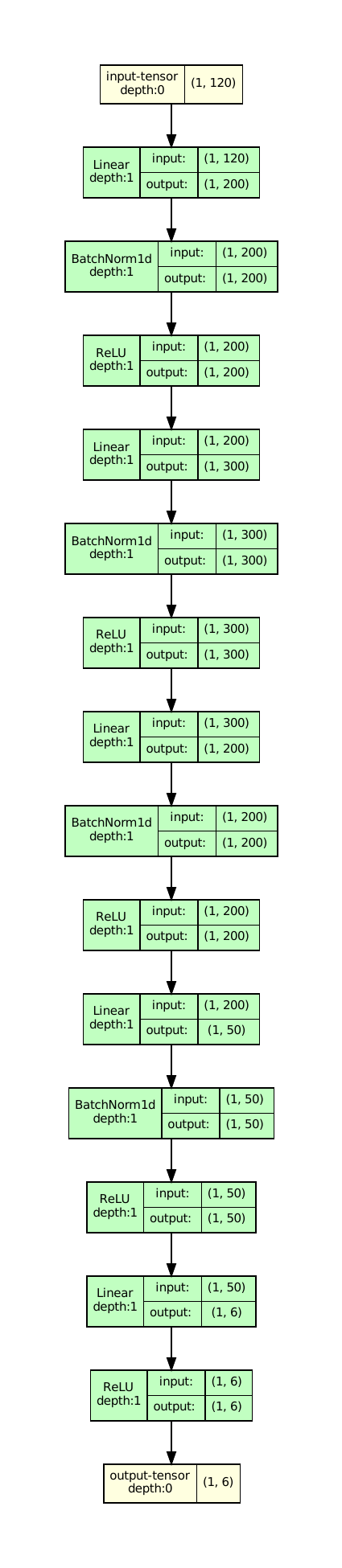}
}
\caption{Machine learning models used for classification for the \texttt{2Fx2T} aperture model. \label{fig:classification-graphs}}
\end{figure*}

\begin{figure*}
\centering
\subfloat[][Disc \label{fig:disc-graph}]
{\includegraphics[width=0.32\textwidth]
{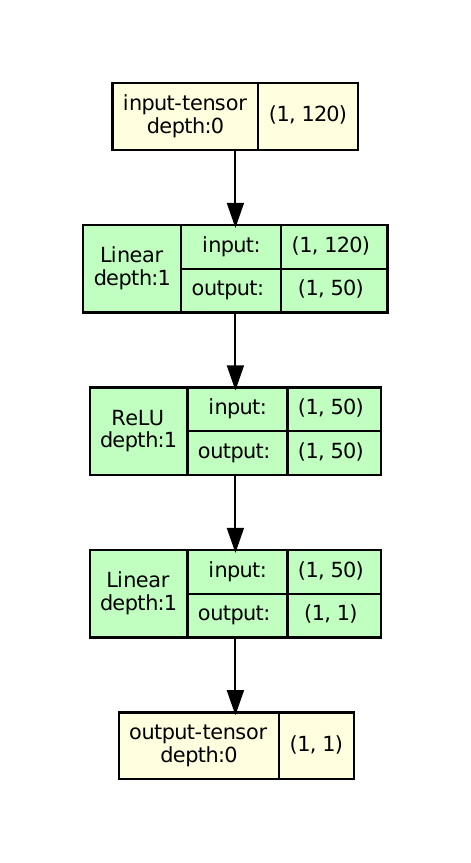}
}
\subfloat[][Crescent \label{fig:crescent-graph}]
{\includegraphics[width=0.34\textwidth,height=0.95\textheight]
{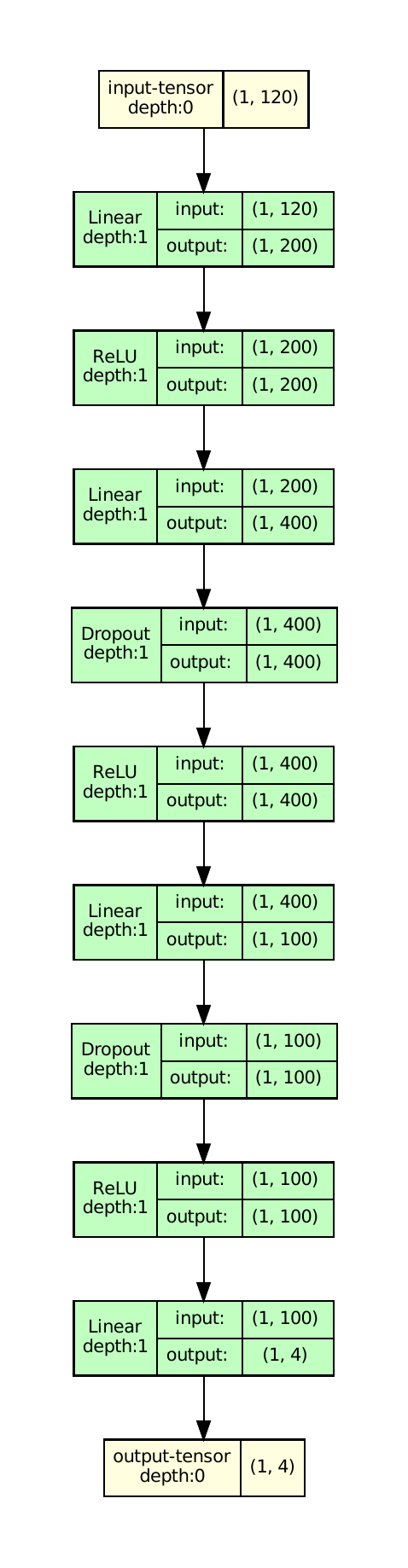}
}
\subfloat[][Dual disc\label{fig:dualdisc-graph}]
{\includegraphics[width=0.34\textwidth,height=0.95\textheight]
{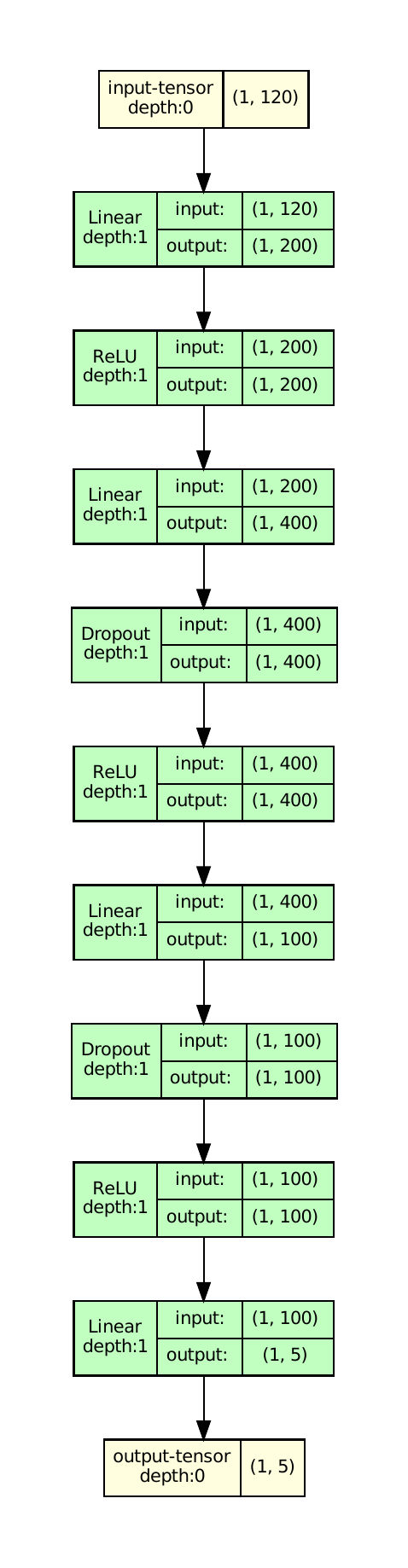}
}
\caption{MLP models used for parametrisation of the \textit{(a)} disc, \textit{(b)} crescent, and \textit{(c)} dual disc morphologies for the \texttt{2Fx2T} aperture model. \label{fig:parametrisation-graphs}}
\end{figure*}

\section{Classification accuracy on noisy data}\label{sec:noiseless-on-noise}

Figure~\ref{fig:F1-noiseless-on-noise} shows the noise performance of our ML classifiers trained on noiseless inputs but tested on noisy inputs, thus representing a worst-case scenario.

\begin{figure}
\centering
\includegraphics[width=\linewidth]
{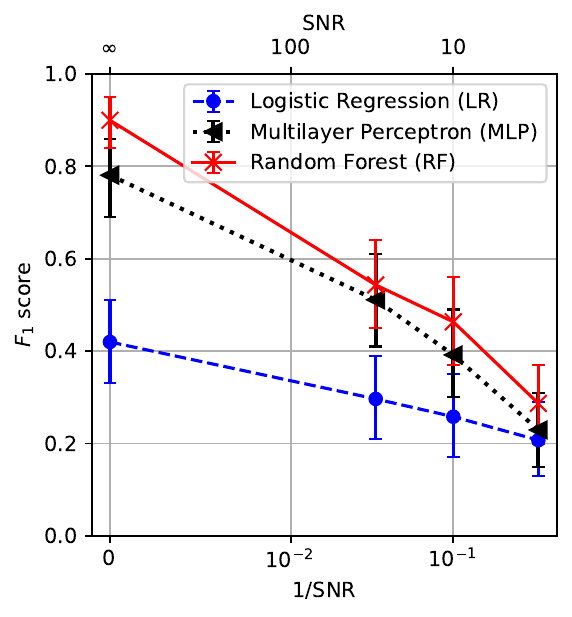}
\caption{Same as Fig.~\ref{fig:F1-noise} but for ML models trained on noiseless data and tested on noisy inputs, representing a worst-case scenario.
  Contrasting with Fig.~\ref{fig:F1-noise}, the accuracy degrades relatively quickly with decreasing signal-to-noise ratio. The error bars denote 95\% confidence intervals. The $x$-axis uses a linear-logarithmic scaling.}
\label{fig:F1-noiseless-on-noise}
\end{figure}


\bsp	
\label{lastpage}
\end{document}